\def \tr {{\rm tr}\,}
\documentclass[showpacs,preprintnumbers,amsmath,amssymb,11pt]{revtex4}
\usepackage{graphicx}

\begin{document}

\title{Towards a gauge-polyvalent Numerical Relativity code}

\author{Daniela Alic, Carles Bona and  Carles Bona-Casas}

\affiliation{Departament de Fisica, Universitat de les Illes
Balears, Palma de Mallorca, Spain.\\
Institute for Applied Computation with Community Code (IAC$\,^3$)}


\begin{abstract}
The gauge polyvalence of a new numerical code is tested, both in
harmonic-coordinate simulations (gauge-waves testbed) and in
singularity-avoiding coordinates (simple Black-Hole simulations,
either with or without shift). The code is built upon an adjusted
first-order flux-conservative version of the Z4 formalism and a
recently proposed family of robust finite-difference
high-resolution algorithms. An outstanding result is the long-term
evolution (up to $1000M\,$) of a Black-Hole in normal coordinates
(zero shift) without excision.
\end{abstract}
\maketitle

\section{Introduction}
In a recent paper~\cite{KS08}, Kiuchi and Shinkai have analyzed
numerically the behavior of many 'adjusted' versions of the BSSN
system. This is a follow-up of a former proposal~\cite{YS02} for
using the energy-momentum constraints to modify Numerical
Relativity evolution formalisms. An important point was to put the
constraint propagation system (subsidiary system) in a strongly
hyperbolic form, so that constraint violations can propagate out
of the computational domain. As a further step, there is also the
possibility of introducing damping terms, which would attract the
numerical solution towards the constrained subspace.

At the first sight, one could wonder why this idea is still
deserving some interest today, when the BSSN system is being
successfully used in binary-black-holes simulations. Waveform
templates are currently being extracted for different mass and
spin configurations, with an accuracy level that depends just on
the computational resources (including the use of mesh-refinement
and/or higher-order finite-difference algorithms). The same is
true for neutron stars simulations, where the BSSN formalism is
currently used for evolving the spacetime
geometry~\cite{NSref1}-\cite{NSref4}. But these success scenarios
have a weak point: the BSSN simulations are based on the
combination of the '1+log' and 'Gamma-driver' gauge conditions, as
proposed in Ref.~\cite{Golm02} for the first long-term dynamical
simulation of a single Black Hole (BH) without excision.

Concerning BH simulations, we can understand that dealing
numerically with collapse singularities requires the use of either
excision, or time slicing prescriptions with strong
singularity-avoidance properties. In the '1+log' case, there is
actually a 'limit hypersurface', so that the numerical evolution
gets safely bounded away from collapse singularities. But
singularity-avoidance is a property of the time coordinate, which
should then be independent of the space coordinates prescription.
In the spirit of General Relativity, we should expect a
gauge-polyvalent numerical code to work as well in normal
coordinates (zero shift), even if some specific type of time
slicing condition (lapse choice) is required for BH simulations.
Moreover, this requirement should be extended to other dynamical
choices of the space coordinates. This means that a
gauge-polyvalent numerical code should also work with alternative
shift prescriptions, provided that the proposed choices preserve
the regularity of the congruence of time lines. And this should be
independent of the fact that a freezing of the dynamics is
obtained or not as a result. These considerations apply \textit{'a
fortriori'} to neutron star simulations without any BH in the
final stage, where no singularity is expected to form.

The above proposed gauge-polyvalence requirements, which are in
keeping with the spirit of General Relativity, may seem too
ambitious, allowing for the fact that they are not fulfilled by
current BH codes. But the need for improvement is even more
manifest by looking at the results of the gauge-waves test. This
test consists in evolving Minkowsky spacetime in non-trivial
harmonic coordinates, and was devised for cross-comparing the
numerical codes performance~\cite{Mexico}. In Ref.~\cite{KS08},
the authors assay different adjustments in order to correct the
poor performance of 'plain' BSSN codes, which was previously
reported in Ref.~\cite{JBT06}. They manage to get long-term
evolutions for the small amplitude case ($A=0.01$) with a standard
second-order-accurate numerical algorithm. The same result was
previously achieved by using a fourth-order accurate finite
differences scheme~\cite{ZBCL05}. Even in this case, however, the
results for the medium amplitude case ($A=0.1$) are disappointing.
More details can be found in a more recent cross-comparison
paper~\cite{Mexico2}, where actually a higher benchmark (big
amplitude, $A=0.5$, devised for testing the non-linear regime) is
proposed.

One could argue that the gauge-waves test is not relevant for real
simulations, because periodic boundary conditions do not allow
constraint violations to propagate out of the computational
domain~\cite{JBT06}. In BH simulations, however, constraint
violations arising inside the horizon can not get out, unless all
the characteristic speeds of the subsidiary system are adjusted to
be greater than light speed. As far as this extreme adjustment is
not implemented in the current evolution formalisms, the
gauge-waves test results can be indeed relevant, at least for
non-excision BH codes. As a result, in keeping with the view
expressed in Ref.~\cite{KS08}, we are convinced that either an
improvement of the current BSSN adjustments or any alternative
formulation would be welcome, as far as it could contribute to
widen the gauge-polyvalence of numerical relativity codes.

In this paper we will consider an alternative numerical code
consisting in two main ingredients. The first one is the Z4
strongly-hyperbolic formulation of the field equations~\cite{Z4}.
The original (second order) version needs no adjustment for the
energy and momentum constraints, as far as constraint deviations
propagate with light speed, although some convenient damping terms
have been also proposed~\cite{damping}. We present in Section II a
first-order version, which has been adjusted for the ordering
constraints which arise in the passage from the second-order to
the first-order formalism. Its flux-Conservative implementation is
described in Appendix A. The second ingredient is the recently
developed FDOC algorithm~\cite{JCPpaper}, which is a (unlimited)
finite-difference version of the Osher-Chakrabarthy finite-volume
algorithm~\cite{ICASE}, along the lines sketched in a previous
paper~\cite{FVpaper}. Although this algorithm, detailed in
Appendix B, allows a much higher accuracy, we will restrict
ourselves here to the simple cases of third and fifth-order
accuracy, which have shown an outstanding robustness, confirmed by
standard tests from Computational Fluid Dynamics, including
multidimensional shock interactions~\cite{JCPpaper}.

The results for the gauge-waves test are presented in
section~\ref{GW}, where just a small amount of dissipation,
without any visible dispersion error, shows up after $1000$
crossing times, even for the high amplitude ($A=0.5$) case.
Simulations of a 3D BH in normal coordinates are presented in
section~\ref{normalBH}, where we consider many variants of the
'Bona-Mass\'{o}' singularity-avoidant prescription~\cite{BMSS}. As
expected, the best results for a given resolution are obtained for
the choices with a limit hypersurface far away from the
singularity. For the $f=2/\alpha$ choice, the BH evolves in normal
coordinates at least up to $1000\,M$ in a uniform grid with
logarithmic space coordinates. This is one order of magnitude
greater than the normal-coordinates BSSN result, as reported
in~\cite{Golm02}.

Concerning the shift conditions, we have tested in
Section~\ref{shift} many explicit first-order prescriptions in
single BH simulations. The idea is just to test the
gauge-polyvalence of the code, so no physically motivated
condition has been imposed, apart from the three-covariance of the
shift under arbitrary time-independent coordinate transformations.
Our results confirm that the proposed code is not specially tuned
for normal coordinates (zero shift).

\section{Adjusting the first-order Z4 formalism}\label{formalism}
The Z4 formalism is a covariant extension of the Einstein field
Equations, defined as~\cite{Z4}
\begin{equation}\label{EEE}
  R_{\mu \nu} + \nabla_{\mu} Z_{\nu} + \nabla_{\nu} Z_{\mu} =
  8\; \pi\; (T_{\mu \nu} - \frac{1}{2}\;T\; g_{\mu \nu}).
\end{equation}
The four vector $Z_{\mu}$ is an additional dynamical field, which
evolution equations can be obtained from (\ref{EEE}). The
solutions of the original Einstein\'{}s equations can be recovered
when $Z_{\mu}$ is a Killing vector. In the generic case, the
Killing equation has only the trivial solution $Z_\mu = 0$, so
that true Einstein's solutions can be easily recognized.

The manifestly covariant form (\ref{EEE}) can be translated into
the 3+1 language in the standard way. The covariant four-vector
$Z_\mu$ will be decomposed into its space components $Z_i$ and the
normal time component
\begin{equation}\label{theta}
    \Theta \equiv n_{\mu}\; Z^{\mu} = \alpha\; Z^0
\end{equation}
where $n_\mu$ is the unit normal to the $t=constant$ slices. The
3+1 decomposition of (\ref{EEE}) is given then by~\cite{Z4}
\begin{eqnarray}
\label{dtgamma}
  (\partial_t -{\cal L}_{\beta})~ \gamma_{ij}
  &=& - {2\;\alpha}\;K_{ij}
\\
\nonumber
   (\partial_t - {\cal L}_{\beta})~K_{ij} &=& -\nabla_i\alpha_j
    + \alpha\;   [R_{ij}
    + \nabla_i Z_j+\nabla_j Z_i
\\
\label{dtK}
    &-& 2K^2_{ij}+(tr\,K - 2\, \Theta)\;K_{ij}
    - 8\pi\{S_{ij}-\frac{1}{2}\,(tr\, S -\; \tau)\;\gamma_{ij}\}\;]
\\
\label{dtTheta} (\partial_t -{\cal L}_{\beta})~\Theta &=&
\frac{\alpha}{2}\;
 [R + 2\; \nabla_k Z^k + (tr\; K - 2\; \Theta)\; tr\, K
 -\tr(K^2)  - 2\; Z^k {\alpha}_k/\alpha - 16\pi\tau]
\\
\label{dtZ}
 (\partial_t -{\cal L}_{\beta})~Z_i &=& \alpha\; [\nabla_j\;({K_i}^j
  -{\delta_i}^j \, tr\,K) + \partial_i \Theta
 - 2\; {K_i}^j\; Z_j  -  \Theta\, {{\alpha}_i/ \alpha} - 8\pi S_i]
~.
\end{eqnarray}

The evolution system can be completed by providing suitable
evolution equations for the lapse and shift components.
\begin{equation}
\label{dtgauge}
 \partial_t \alpha =  - \alpha^2~Q~,
\qquad
 \partial_t \beta^i = -~\alpha~Q^i
\end{equation}
We will keep open at this point the choice of gauge conditions, so
that the gauge-derived quantities $\{ Q,\, Q^i \}$ can be either a
combination of the other dynamical fields or independent
quantities with their own evolution equation. We are assuming,
however, that both lapse and shift are dynamical quantities, so
that terms involving derivatives of $\{ Q,\, Q^i \}$ actually
belong to the principal part of the evolution system.

\subsection*{First-order formulation: ordering constraints} %
In order to translate the evolution system
(\ref{dtgamma}-\ref{dtgauge}) into a fully first-order form, the
space derivatives of the metric components (including lapse and
shift) must be introduced as new independent quantities:
\begin{equation}\label{space_derivatives}
 A_i~\equiv~ \partial_i \ln \alpha,
 ~~{B_{k}}^i~\equiv~ \partial_k \beta^{i},
 ~~D_{kij}~\equiv~\frac{1}{2}\;\partial_k \gamma_{ij}~.
\end{equation}

Note that, as far as the new quantities will be computed now
through their own evolution equations, the original definitions
(\ref{space_derivatives}) must be considered rather as constraints
(first-order constraints), namely
\begin{eqnarray}\label{ak_constraint}
  {\cal A}_k \equiv A_k - \partial_k\,ln\alpha &=& 0 \\ \label{Bk_constraint}
  {{\cal B}_k}^i \equiv {B_k}^i - \partial_k\,\beta^i &=& 0\\
  {\cal D}_{kij} \equiv  D_{kij}-\frac{1}{2}\,\partial_k \gamma_{ij}
  &=& 0\,.\label{Dk_constraint}
\end{eqnarray}
Note also that we can derive in this way the following set of
constraints, related with the ordering of second derivatives
(ordering constraints):
\begin{eqnarray}
\label{orderingA}
    {\cal C}_{ij}&\equiv&\partial_i\,{\cal A}_j - \partial_j\,{\cal A}_i
    = \partial_i\,A_j - \partial_j\,A_i~=~0\,, \\
\label{orderingB}
    {{\cal C}_{rs}}^i&\equiv&\partial_r\,{{\cal B}_s}^i
    - \partial_s\,{{\cal B}_r}^i = \partial_r\,{B_s}^i
    - \partial_s\,{B_r}^i = 0\,, \\
\label{orderingD}
    {\cal C}_{rsij}&\equiv&\partial_r\,{\cal D}_{sij} -
    \partial_s\,{\cal D}_{rij} =
    \partial_r\,D_{sij} - \partial_s\,D_{rij}~=~0\,.
\end{eqnarray}

The evolution of the lapse and shift space derivatives could be
obtained easily, just by taking the time derivative of the
definitions (\ref{space_derivatives}) and exchanging the order of
time and space derivatives. But then the characteristic lines for
the transverse-derivative components in (\ref{space_derivatives})
would be the time lines (zero characteristic speed). This can lead
to a characteristic  degeneracy problem, because the
characteristic cones of the second-order system
(\ref{dtK}-\ref{dtZ}) are basically the light cones~\cite{Z4}, and
the time lines can actually cross the light cones, as it is the
case in many black hole simulations. In order to avoid this
degeneracy problem, we can make use of the shift ordering
constraint (\ref{orderingB}) for obtaining the following evolution
equations for the additional quantities (\ref{space_derivatives}):
\begin{eqnarray}
\label{dtA1}
\partial_t A_{k} &+& \partial_l [ -\beta^l~A_{k} +
        {\delta^l}_k~(\alpha~Q + \beta^rA_{r} ) ] = {B_k}^l~A_{l} - tr B~A_{k}
\\
\label{dtB1}
 \partial_t {B_k}^i &+& \partial_l [ -\beta^l~{B_k}^i +
        {\delta^l}_k~(\alpha~Q^i + \beta^r{{B_r}^i} ) ] = {B_k}^l~{B_l}^i - tr B~{B_k}^i
\\
\label{dtD1}
 \partial_t D_{kij} &+& \partial_l [ -\beta^lD_{kij} +
        {\delta^l}_k~\{\alpha~K_{ij} - 1/2~(B_{ij}+B_{j\,i})\}\,] = {B_k}^l~D_{lij} -
        tr B~D_{kij}~.
\end{eqnarray}

Note that the characteristic lines for the transverse-derivative
components are now the normal lines (instead of the time lines),
so that characteristic crossing is actually avoided. This ordering
adjustment is crucial for long-term evolution in the dynamical
shift case, as it has been yet realized in the first-order version
of the generalized harmonic formulation~\cite{HLOPSK04}.

\subsection*{Damping terms adjustments}

A further adjustment could be the introduction of some
constraint-violation damping terms. For the energy-momentum
constraints, these terms can be added to the evolution equations
(\ref{dtK}-\ref{dtZ}), as described in Ref.~\cite{damping}.

For the ordering constraints, we can also introduce simple
constraint-violation damping terms when required. For instance,
equation (\ref{dtA1}) could be modified as follows:
\begin{equation}\label{Adamped}
\partial_t A_{i} + \partial_l [ -\beta^lA_{i} +
        {\delta^l}_i~(\alpha~Q + \beta^rA_{r} ) ] = {B_i}^l~A_{l} - tr B~A_{i}
        - \eta~ {\cal A}_{i}~,
\end{equation}
with the damping parameter in the range $0\leq\eta\ll 1/\Delta t$.
The same pattern could be applied to equations (\ref{dtB1},
\ref{dtD1}).

In order to justify this, let us analyze the resulting evolution
equations for the first-order constraints (\ref{ak_constraint}).
Allowing for (\ref{dtA1}), we would get
\begin{equation}\label{Ak_subs}
    \partial_t\,{\cal A}_{k}
    - \beta^r\,(\partial_r\,{\cal A}_{k}
    -\partial_k\,{\cal A}_{r}) = {{\cal B}_k}^r\,A_{r}
    -{{\cal B}_r}^r\,A_{k}\,.
\end{equation}\index{Subsidiary system}
The hyperbolicity of the subsidiary evolution equation
(\ref{Ak_subs}) can be analyzed by looking at the normal and
transverse components of the principal part along any space
direction $\vec{n}$, namely
\begin{eqnarray}\label{An_subs}
  \partial_t\,{\cal A}_{n}
    - \beta^\bot\,(\partial_n\,{\cal A}_{\bot}) &=& 0\\
 \partial_t\,{\cal A}_{\bot}
    - \beta^n\,(\partial_n\,{\cal A}_{\bot})
     &=& 0\,,\label{Atrans_subs}
\end{eqnarray}
with eigenvalues ($0$, $- \beta^n$), which is just weakly
hyperbolic in the fully degenerate case, that is for any space
direction orthogonal to the shift vector. Note that this is just
the subsidiary system governing constraint violations, not the
evolution system itself. This means that the main concern here is
accuracy, rather than stability. But the resulting (linear)
secular growth of first-order constraint violations may become
unacceptable in long-term simulations.

These considerations explain the importance of adding
constraint-damping terms, so that (\ref{dtA1}) is replaced by
(\ref{Adamped}). The damping term $-\eta\,{\cal A}_{k}$ will
appear as a result in the subsidiary system  also. The linearly
growing constraint-violation modes arising from the degenerate
coupling in (\ref{An_subs}) will be kept then under control by
these (exponential) damping terms. The same argument applies {\em
mutatis mutandis} to the remaining first-order constraints ${{\cal
B}_k}^i$, ${\cal D}_{kij}\,$.

\subsection*{Secondary ordering ambiguities}

The shift ordering constraints (\ref{orderingB}) can also be used
for modifying the first-order version of the evolution equation
(\ref{dtZ}) in the following way
\begin{equation}\label{dtZ_adjusted}
 (\partial_t -{\cal L}_{\beta})~Z_i = \alpha\, [\,\nabla_j\,({K_i}^j
  -{\delta_i}^j \, tr\,K) + \partial_i\, \Theta
 - 2\, {K_i}^j\, Z_j  -  \Theta\, A_i - 8\pi S_i\,]
 - \mu~(\partial_j \,{B_{i}}^j - \partial_i\, tr B)
\,.
\end{equation}
Also, the ordering constraints (\ref{orderingD}) can be used for
selecting a specific first-order form for the three-dimensional
Ricci tensor appearing in (\ref{dtK})~\cite{Z48}. This can be any
combination of the standard Ricci decomposition
\begin{equation}\label{Def3R}
R_{ij}~=~\partial_k\,{\Gamma^k}_{ij}-\partial_{\,i}\,{\Gamma^k}_{kj}
+{\Gamma^r}_{rk}\,{\Gamma^k}_{ij}-{\Gamma^k}_{ri}\,{\Gamma^r}_{kj}
\end{equation}
with the De~Donder  decomposition
\begin{eqnarray}\label{Def3dD}
R_{ij}&=&-\partial_k\,{D^k}_{ij}+\partial_{(i}\,{\Gamma_{j\,)k}}^{k}
- 2 {D_r}^{rk} D_{kij} \nonumber \\
&+& 4\, {D^{rs}}_i D_{rsj} - {\Gamma_{irs}}
{\Gamma_j\,}^{rs}-{\Gamma_{rij}}\, {\Gamma^{rk}}_k
\end{eqnarray}
which is most commonly used in Numerical Relativity codes.
Following Ref.~\cite{Z48}, we will introduce an ordering parameter
$\xi$, so that $\xi=1$ corresponds to the Ricci decomposition
(\ref{Def3R}) and $\xi=-1$ to the De~Donder one (\ref{Def3dD}).

The choices of $\mu$ and $\xi$ do not affect the characteristic
speeds of the evolution system (see Appendix A for details), nor
the structure of the subsidiary system. In this sense, these are
rather secondary ordering ambiguities and we will keep these
parameters free for the moment, although there are some
prescriptions that can be theoretically motivated:
\begin{itemize}
    \item The choice $\mu=1/2$,\, $\xi=-1$ allows to recover at the
first-order level the equivalence between the generalized harmonic
formulation and (the second-order version of) the Z4 formalism,
given by~\cite{damping}
\begin{equation}\label{equivalence}
    Z^\mu =
    \frac{1}{2}\,{\Gamma^{\mu}}_{\rho\sigma}\,g^{\rho\sigma}\,
\end{equation}
(see Appendix A for more details). This can be important, because
the harmonic system is known to be symmetric hyperbolic.
    \item The choice $\mu=1$ is the only one that ensures the
    strong hyperbolicity of the Z3 system, obtained from the Z4
    one by setting $\theta=0$. This can be relevant if we are
    trying to keep energy-constraint violations close to zero.
    Allowing for the quasi-equivalence between the Z3 and the BSSN
    systems~\cite{Z48}, this adjustment will affect as well to the
    first-order version of the BSSN system (NOR system~\cite{NOR})
    in simulations using dynamical shift conditions. The same
    comment applies to the old 'Bona-Mass{\' o}' system~\cite{BM92}.
    \item The choice $\xi=0$ ensures that the first-order version
    contains only symmetric
    combinations of second derivatives of the space metric. This
    is a standard symmetrization procedure for obtaining a first-order
    version of a generic second-order equation.
\end{itemize}
In the numerical simulations in this paper, we have taken
$\mu=1$,\, $\xi=-1$, although we have also tested other
combinations, which also lead to long-term stability.

\section{Gauge waves test}\label{GW}
We will begin with a test devised for harmonic coordinates. Let us
consider the following line element:
\begin{equation}\label{gaugew}
    ds^2 = H(x-t)(-dt^2 + dx^2) + dy^2 + dz^2 ~,
\end{equation}
where $H$ is an arbitrary function of its argument. One could
naively interpret this as the propagation of an arbitrary wave
profile with unit speed. But it is a pure gauge effect, because
(\ref{gaugew}) is nothing but the Minkowsky metric, written in
some non-trivial harmonic coordinates system.

As proposed in Refs.~\cite{Mexico},~\cite{Mexico2}, we will
consider the 'gauge waves' line element (\ref{gaugew}), with the
following profile:
\begin{equation}\label{sinusw}
    H = 1 - A~Sin(~2\pi (x-t)~)\,,
\end{equation}
so that the resulting metric is periodic and we can identify for
instance the points $-0.5$ and $0.5$ on the $x$ axis. This allows
to set up periodic boundary conditions in numerical simulations,
so that the initial profile keeps turning around along the $x$
direction. One can in this way test the long term effect of these
gauge perturbations. The results show that the linear regime
(small amplitude, $A=0.01$) poses no serious challenge to most
Numerical Relativity codes (but see Ref.~\cite{KS08} for the BSSN
case). Following the recent suggestion in Ref.~\cite{Mexico2}, we
will then focus in the medium and big amplitude cases ($A=0.1$ and
$A=0.5$, respectively), in order to test the non-linear regime.
Concerning grid spacing, although $\Delta x= 0.01$ would be enough
for passing the test in the medium amplitude case, the big
amplitude one requires more resolution, so we have taken $\Delta
x= 0.005$ in both cases.

\begin{figure}[t]
\centering \scalebox{0.5}[0.6]{\includegraphics{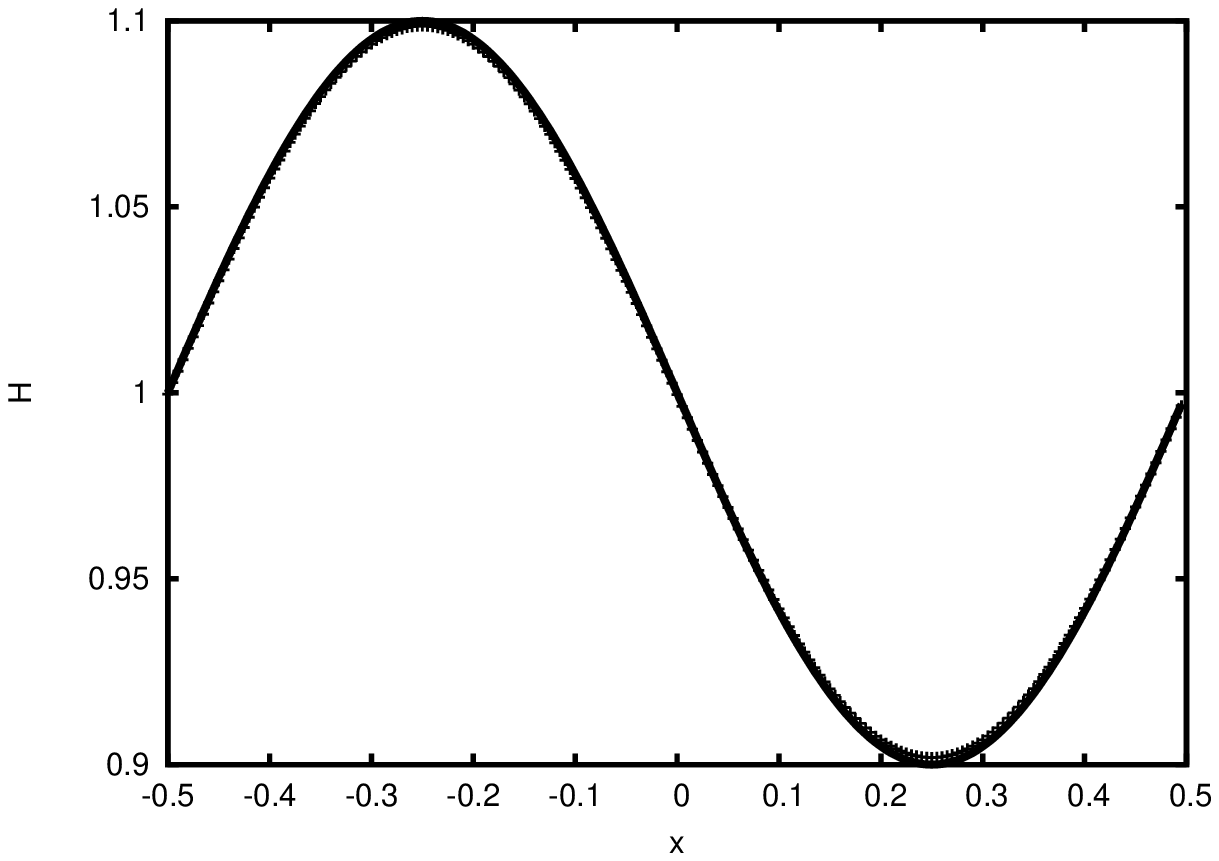}
\includegraphics{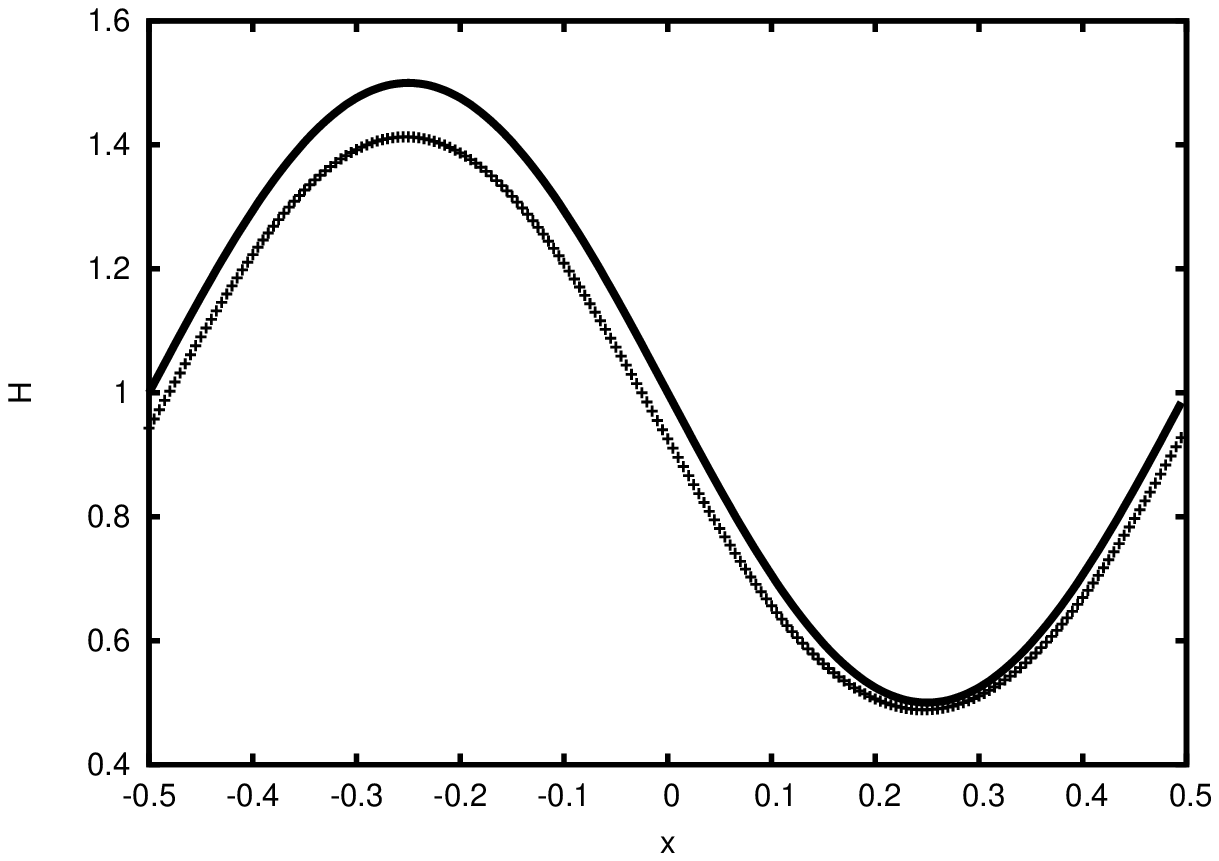}}
\caption{Gauge waves simulation with periodic boundary conditions
and sinusoidal initial data for the $\gamma_{xx}$ metric
component. The resolution is $\Delta x= 0.005$ in both cases. The
left panel corresponds to the medium amplitude case $A=0.1$. After
$1000$ round trips, the evolved profile (cross marks) nearly
overlaps the initial one (continuous line), which corresponds also
with the exact solution. The right panel corresponds to the same
simulation for the big amplitude case $A=0.5$. We see the
combination of a slight decrease in the mean value plus some
amplitude damping.} \label{gaugewplots}
\end{figure}

The results of the numerical simulations are displayed in
Fig.~\ref{gaugewplots} for the $H$ function (the $\gamma_{xx}$
metric component). The left panel shows the medium amplitude case
$A=0.1$. Only a small amount of numerical dissipation is barely
visible after $1000$ round trips: the third-order-accurate
finite-difference method gets rid of the dominant dispersion
error. For comparison, let us recall that the corresponding BSSN
simulation crashes before $100$ round trips~\cite{ZBCL05}. The
right panel shows the same thing for the large amplitude case
$A=0.5$, well inside the non-linear regime. We see some amplitude
damping, together with a slight decrease of the mean value of the
lapse.

Our results are at the same quality level than the ones reported
in Ref.~\cite{Mexico2} for the Flux-Conservative
generalized-harmonic code Abigail (see also the 'apples with
apples' webpage ~\cite{Abigail}), which is remarkable for a test
running in strictly harmonic coordinates. We can also compare with
the simulations reported in Ref.~\cite{TLN} for (a specific
variant of) the KST evolution system~\cite{KST}. Although the
gauge wave parametrization is not the standard one, both their
'big amplitude' case and their finest resolution are similar to
ours. We see a clear phase shift, due to cumulative dispersion
errors, after about $500$ crossing times. We see also a growing
amplitude mode, which can be moderated with resolution (for the
finest one, it just compensates numerical dissipation). This can
be related with the spurious linear mode that has been reported
for harmonic systems which are not written in Flux-Conservative
form~\cite{Mexico}.

We can conclude that there are two specific ingredients in our
code that contribute to the gauge-wave results in an essential
way: the Flux-Conservative form of the equations (see Appendix A),
which gets rid of the spurious growing amplitude modes, and the
third-order accuracy of the numerical algorithm, which reduces the
dispersion error below the visual detection level in
Fig.~\ref{gaugewplots}, even after $1000$ crossing times.

\section{Single Black hole test: normal coordinates}\label{normalBH}
We will try next to test a Schwarzschild black-hole evolution in
normal coordinates (zero shift). Harmonic codes are not devised
for this gauge choice, so we will compare with BSSN results
instead. Concerning the time coordinate condition, our choice will
be limited by the singularity-avoidance requirement, as far as we
are not going to excise the black-hole interior. Allowing for
these considerations, we will determine the gauge evolution
equations (\ref{dtgauge}) as follows
\begin{equation}\label{fgauge}
    Q = f~(tr\,K-m\,\Theta)~, \qquad Q^i=0~~(\beta^i = 0)~,
\end{equation}
where the second gauge parameter $m$ is a feature of the Z4
formalism. We will choose here by default $m=2$, because the
evolution equation for the combination $trK-2\,\Theta$, as derived
from (\ref{dtK}, \ref{dtTheta}), actually corresponds with the
BSSN evolution equation for $tr\,K$ (see Ref.~\cite{Z48} for the
relationship between BSSN and Z4 formalisms).

Concerning the first gauge parameter, we will consider first the
'1+log' choice $f=2/\alpha$~\cite{Bernstein}, which is the one
used in current binary BH simulations in the BSSN formalism. The
name comes from the resulting form of the lapse, after integrating
the evolution equation (\ref{dtgamma}, \ref{dtgauge}) with the
prescription (\ref{fgauge}) for true Einstein's solutions
($\Theta=0$):
\begin{equation}\label{1+log}
    \alpha = \alpha_0 + ln\,(\gamma/\gamma_0)~,
\end{equation}
where $\sqrt{\gamma}$ is the space volume element. It follows from
(\ref{1+log}) that the coordinate time evolution stops at some
limit hypersurface, before even getting close to the collapse
singularity. This happens when
\begin{equation}\label{limitsurf}
    \sqrt{\gamma/\gamma_0} = exp\,(-\alpha_0/2)~,
\end{equation}
that is well before the vanishing of the space volume element: the
initial lapse value is usually close to one, so that the final
volume element is still about a $60\%$ of the initial one. This
can explain the robustness of the 1+log choice in current
black-hole simulations.

We will consider as usual initial data on a time-symmetric time
slice ($K_{ij}=0$) with the intrinsic metric given in isotropic
coordinates:
\begin{equation}\label{isotropic}
    \gamma_{ij} = (1+\frac{m}{2r})^4~\delta_{ij}~.
\end{equation}
This is the usual 'puncture' metric, with the apparent horizon at
$r=m/2$: the interior region is isometric to the exterior one, so
that the $r=0$ singularity is actually the image of space
infinity. We prefer, however, to deal with non-singular initial
data. We will then replace the constant mass profile in interior
region $r<M/2$ by a suitable profile $m(r)$, so that the interior
metric corresponds to a scalar field matter content. Of course,
the scalar field itself must be evolved consistently there (see
Appendix C for details). A previous implementation of the same
idea, with dust interior metrics, can be found in
Ref.~\cite{stuffing}.

\begin{figure}[h]
\centering {
\includegraphics{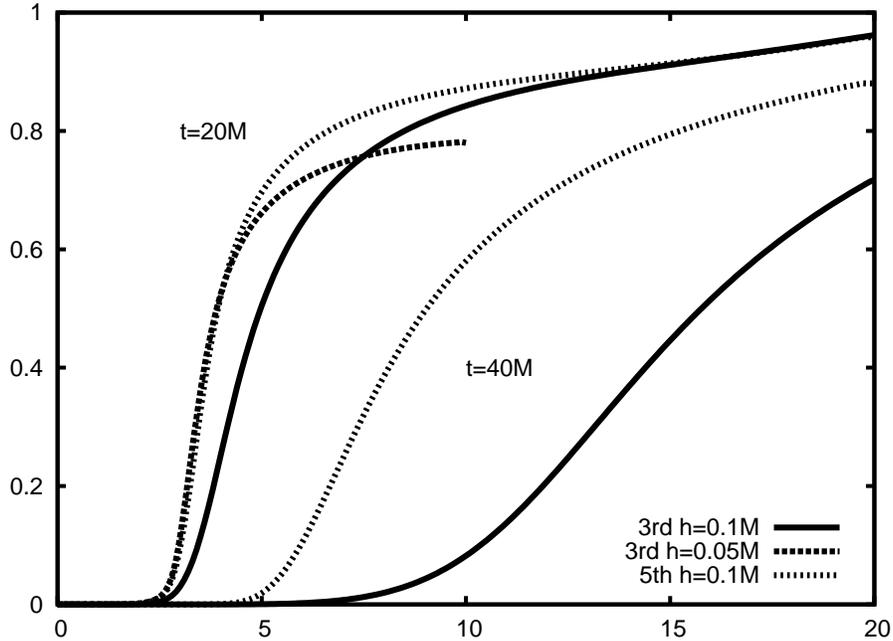}
} \caption{Plots of the lapse profiles at $t=20M\,$ and $t=40\,M$.
The results for the third-order accurate algorithm (continuous
lines) are compared with those for the fifth-order algorithm
(dotted lines) for the same resolution ($h=0.1M\,$). We have also
included for comparison one extra line, corresponding to the
third-order results with $h=0.05\,M$, computed in a reduced mesh.
Increasing resolution leads to a slope steepening and a slower
propagation of the collapse front. In this sense, as we can see
for $t=20\,M$, switching to the fifth-order algorithm while
keeping $h=0.1\,M$ amounts to doubling the resolution for the
third-order algorithm.}\label{BHshort}
\end{figure}

We have performed a  numerical simulation for the $f=2/\alpha$
case with a uniform grid with resolution $h=0.1\,M$, extending up
to $r=20\,M$ (no mesh-refinement). We have used the third and
fifth-order FDOC algorithms, as described in Appendix B, with the
optimal dissipation parameters for each case. The results for the
lapse profile are shown in Fig.~\ref{BHshort} at $t=20\,M$ an
$t=40\,M$. We see in both cases that the higher order algorithm
leads to steeper profiles and a slower propagation of the collapse
front. Note that the differences in the front propagation speed
keep growing in time, although the third-order plot at $t=40\,M$
is clearly affected by the vicinity of the outer boundary. This
fact does not affect the code stability, as far as we can proceed
with the simulations beyond $t=50\,M$, when the collapse front
gets out of the computational domain (beyond $t=60\,M$ in the
higher-order simulations). Note that the corresponding BSSN
simulations ($f= 2/\alpha$ in normal coordinates) are reported to
crash at about $t=40\,M$~\cite{Golm02}.

We have added for comparison an extra plot in Fig.~\ref{BHshort},
with the results at $t=20\,M$ of a third-order simulation with
double resolution ($h=0.05\,M$), obtained in a smaller
computational domain (extending up to $10\,M$). Both the position
and the slope of the collapse front coincide with those of the
fitfth-order algorithm with $h=0.1\,M$. In this case, switching to
the higher-order algorithm amounts to doubling accuracy. Note,
however, that higher-order algorithms are known to be less
robust~\cite{JCPpaper}. Moreover, as the profiles steepen, the
risk of under-resolution at the collapse front increases. We have
found that a fifth-order algorithm is a convenient trade-off for
our $h=0.1\,M$ resolution in isotropic coordinates.

We have also explored other slicing prescriptions with limit
surfaces closer to the singularity, as described in
Table~\ref{table}. Note that in these cases the collapse front
gets steeper than the one shown in Fig.~\ref{BHshort} for the
standard $f=2/\alpha$ case with the same resolution. This poses an
extra challenge to numerical algorithms, so we have switched to
the third-order-accurate one for the sake of robustness. In all
cases, the simulations reached $t=50\,M$ without problem, meaning
that the collapse front has get out of the computational domain.
It follows that the standard prescription $f=2/\alpha$, although
it leads actually to smoother profiles, is not crucial for code
stability.

\begin{table}[h]
  \centering
  \begin{tabular}{|c|c|c|c|c|}
    \hline
    f &\, 2/$\alpha$ &\,1+1/$\alpha$&\, 1/2+1/$\alpha$
    &\, 1/$\alpha$\\
    \hline
    $\,\sqrt{\gamma/\gamma_0}\,$ &\, 61\% & 50\% & 44\% &\, 37\%  \\
    \hline
  \end{tabular}
\caption{Different prescriptions for the gauge parameter
  $f$, with the corresponding values of the  residual volume element
  at the limit surface (normal coordinates), assuming a unit value of the
  initial lapse.\label{table}}
\end{table}

The results shown in Fig.~\ref{BHshort} compare with the ones in
Ref.~\cite{Robust}, obtained with (a second-order version of) the
old Bona-Mass{\'o} formalism. We see the same kind of steep
profiles, produced by the well known slice-stretching
mechanism~\cite{stretching}. This poses a challenge to standard
numerical methods: in Ref.~\cite{Robust} Finite-Volume methods
where used, including slope limiters. Our FDOC algorithm (see
Ref.~\cite{JCPpaper} for details) can also be interpreted as an
efficient Finite-Differences (unlimited) version of the
Osher-Chakrabarthy Finite-Volume algorithm~\cite{ICASE}. Note
however that in Ref.~\cite{Robust}, like in the BSSN case, a
conformal decomposition of the space metric was considered, and an
spurious (numerical) trace mode arise in the trace-free part of
the extrinsic curvature. An additional mechanism for resetting
this trace to zero was actually required for stability. In our
(first-order) Z4 simulations, both the plain space metric and
extrinsic curvature can be used directly instead, without
requiring any such trace-cleaning mechanisms.

\begin{figure}[h]
\centering {\includegraphics{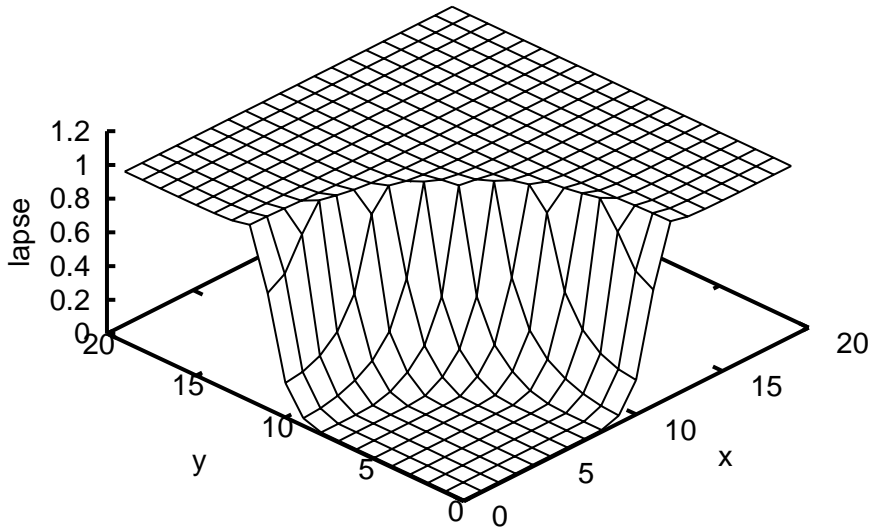}} \caption{Plot of the
lapse function for a single BH at $t=1000\,M$ in normal
coordinates. Only one of every ten points is shown along each
direction. The third-order accurate algorithm has been used with
$\beta=1/12$ and a space resolution $h=0.1\,M$. The profile is
steep, but smooth: no sign of instability appears. Small riddles,
barely visible on the top of the collapse front, signal some lack
of resolution because of the logarithmic character of the grid.
The dynamical zone is safely away from the
boundaries.}\label{BHlong}
\end{figure}

Let us take one further step. Note that the lifetime of our
isotropic coordinates simulations (with no shift) is clearly
limited by the vicinity of the boundary (at $r=20\,M$). At this
point, we can appeal to space coordinates freedom, switching to
some logarithmic coordinates, as defined by
\begin{equation}\label{logcoords}
    R = L~sinh(r/L)~,
\end{equation}
where $R$ is the new radial coordinate and $L$ some length scale
factor. This configuration suggests using the third-order
algorithm because of its higher robustness. We have performed a
long-term numerical simulation for the $f=2/\alpha$ case, with
$L=1.5\,M$, so that $R=20\,M$ in these logarithmic coordinates
corresponds to about $r=463.000\,M$ in the original isotropic
coordinates. In this way, as shown in Fig.~\ref{BHlong}, the
collapse front is safely away from the boundary, even at very late
times. We stopped our code at $t=1000\,M$, without any sign of
instability. This provides a new benchmark for Numerical
Relativity codes: a long-term simulation of a single black-hole,
without excision, in normal coordinates (zero shift). Moreover, it
shows that a non-trivial shift prescription is not a requisite for
code stability in BH simulations.

\section{Single Black hole test: first-order shift
conditions}\label{shift} Looking at the results of the previous
Section, one can wonder wether our code is just tuned for normal
coordinates. This is why we will consider here again BH
simulations, but this time with some non-trivial shift
prescriptions. The idea is just to test some simple cases in order
to show the gauge-polyvalence of the code. For the sake of
simplicity, we will consider here just first order shift
prescriptions, meaning that the source terms ($Q$, $Q^i$) in the
gauge evolutions (\ref{dtgauge}) are algebraic combinations of the
remaining dynamical fields. To be more specific, we shall keep
considering slicing conditions defined by
\begin{equation}\label{fgauge_shift}
    Q = -\beta^k/\alpha\,A_k + f~(tr\,K-m\,\Theta)~,
\end{equation}
together with dynamical shift prescriptions, defined by different
choices of $Q^i$.

First-order shift prescriptions have been yet considered at the
theoretical level~\cite{shift}. We will introduce here an
additional requirement, which follows when realizing that,
allowing for the 3+1 decomposition of the line element
\begin{equation}\label{ds2}
    ds^2 = -\alpha^2~dt^2 +
    \gamma_{ij}\,(dx^i+\beta^idt)\,(dx^j+\beta^jdt)~,
\end{equation}
the shift behaves as a vector under (time independent)
transformations of the space coordinates. We will impose then that
its evolution equation, and then $Q^i$, is also three-covariant.

This three-covariance requirement could seem a trivial one. But
note that the harmonic shift conditions, derived from
\begin{equation}\label{boxx}
    \Box~x^i = 0,
\end{equation}
are not three-covariant (the box here stands for the wave operator
acting on scalars). In the 3+1 language, (\ref{boxx}) can be
translated as
\begin{equation}\label{harmonic_shift}
    \partial_t (\sqrt{\gamma}/\alpha~\beta^i) -
    \partial_k (\sqrt{\gamma}/\alpha~\beta^k\beta^i) +
    \partial_k (\alpha\sqrt{\gamma}~ \gamma^{ik}) = 0~,
\end{equation}
where the non-covariance comes from the space-derivatives terms.

Concerning the advection term, a three-covariant alternative would
be provided either by the Lie-derivative term
\begin{equation}\label{lieterm}
    {\cal L}_\beta\,(\,\sqrt{\gamma}~\beta^{i}/\alpha\,) =
    {\cal L}_\beta\,(\,\sqrt{\gamma}/\alpha\,)~\beta^{i}~,
\end{equation}
or by the three-covariant derivative term
\begin{equation}\label{covderiv_term}
    \beta^k\,\nabla_k (\,\beta^i/\alpha\,) =
    1/\alpha\,[\,\beta^k\,{B_k}^i- \beta^i\beta^kA_k
    +\Gamma^i_{j\,k}\,\beta^j\beta^k\,]~.
\end{equation}
We have tested both cases in our numerical simulations.

Concerning the last term in (\ref{harmonic_shift}), we can take
any combination of $A^i$, $Z^i$ and the vectors obtained form the
space metric derivatives after subtracting their initial values,
namely:
\begin{equation}
  D_i - D_i\,|_{t=0}~, \qquad
  E_i - E_i\,|_{t=0}~.
\end{equation}
This is because the additional terms arising in the transformation
of the non-covariant quantities ($D_i$, $E_i$) depend only on the
space coordinates transformation, which is assumed to be
time-independent. Note that, for the conformal contracted-Gamma
combination
\begin{equation}\label{gamma}
    \Gamma_i = 2\,E_i - \frac{2}{3}~D_i~,
\end{equation}
the subtracted terms actually vanish in simulations starting from
the isotropic initial metric (\ref{isotropic}). Of course, the
same remark applies to the BSSN Gamma quantity, namely~\cite{Z48}
\begin{equation}\label{gamma_BSSN}
    \widetilde{\Gamma}_i = \Gamma_i + 2\,Z_i~.
\end{equation}

We have considered the following combinations:
\begin{eqnarray}\label{S1}
  S1: &~& \partial_t\,\beta^i = \frac{\alpha^2}{2}~A^i
    -\alpha\,Q\,\beta^i \\ \label{S2}
  S2: &~& \partial_t\,\beta^i = \frac{\alpha^2}{2}~A^i
    + \beta^k\,{B_k}^i +\Gamma^i_{j\,k}\,\beta^j\beta^k -\alpha\,Q\,\beta^i \\
  S3: &~& \partial_t\,\beta^i = \frac{\alpha^2}{4}~\widetilde{\Gamma}^i
            + \beta^k\,{B_k}^i +\Gamma^i_{j\,k}\,\beta^j\beta^k
            -\alpha\, Q\,\beta^i ~,\label{S3}
\end{eqnarray}
where S1 corresponds to the Lie-derivative term (\ref{lieterm})
and the remaining two choices to the covariant advection term
(\ref{covderiv_term}), with different combinations of the
first-order vector fields.

\begin{figure}[h]
\centering
\scalebox{0.5}[0.6]{
\includegraphics{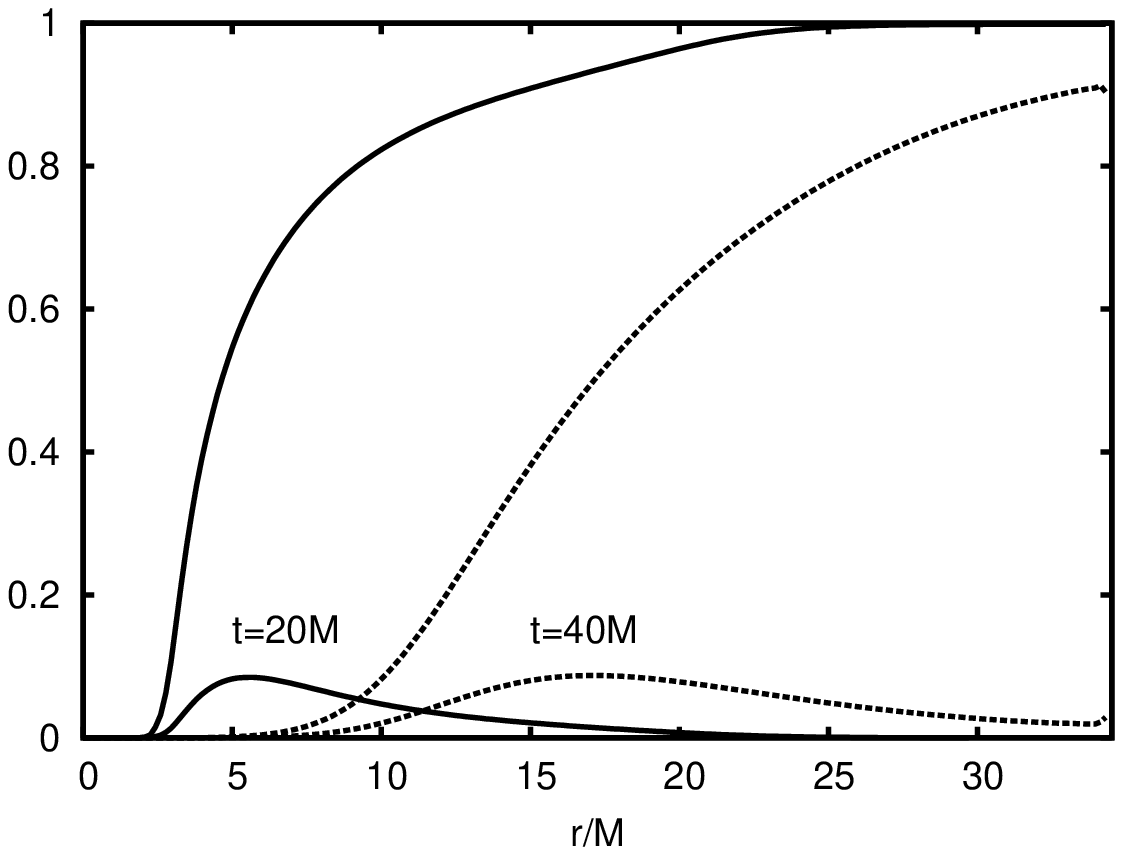}
\includegraphics{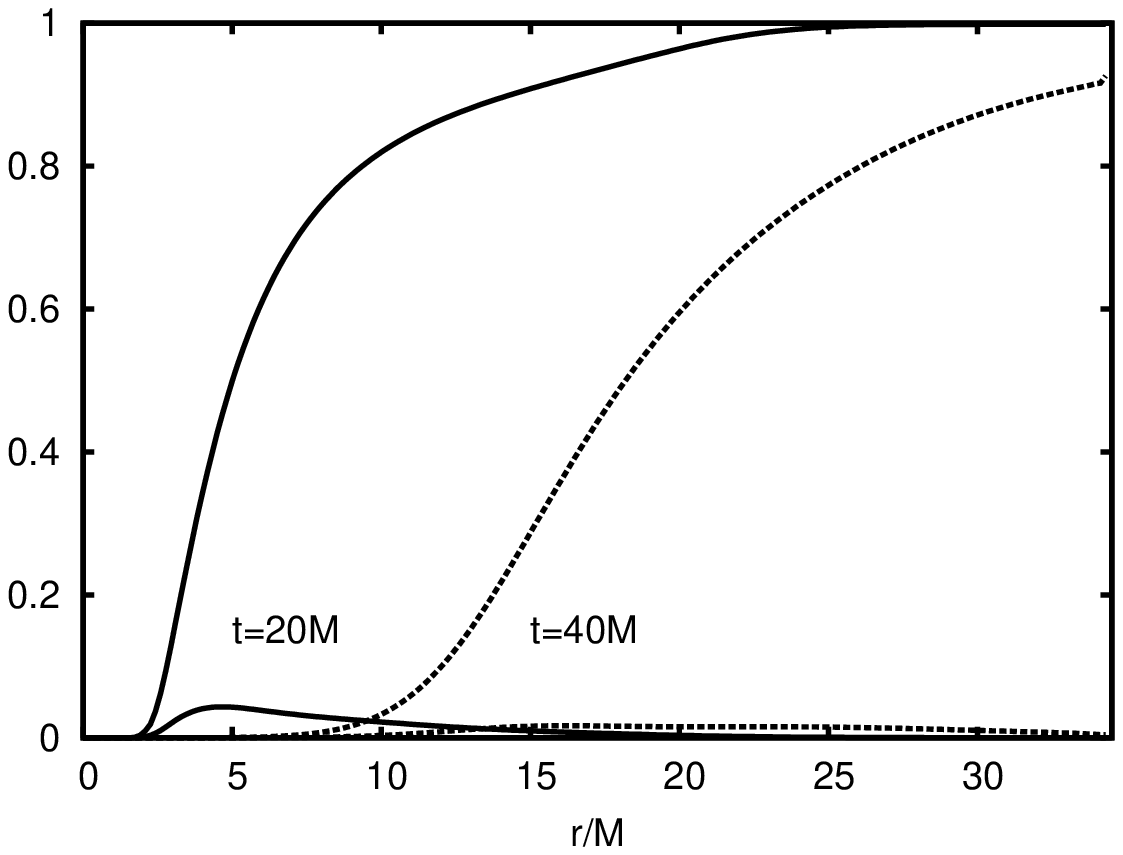}}
\caption{Plot of the lapse and shift profiles at $t=20\,M$
(continuous lines) and $t=40\,M$ (dotted lines). The plots are
shown along the main diagonal of the computational domain, in
order to keep the outer boundary out of the dynamical zone. In the
S1 case (left panel), after the initial growing, the maximum shift
value keeps constant. In the S3 case (right panel), it clearly
diminishes with time.}\label{shiftplots}
\end{figure}

We have obtained stable evolution in all cases, with the
simulations lasting up to the point when the collapse front
crosses the outer boundary (about $t=50\,M$). We can see in
Fig.~\ref{shiftplots} the lapse and shift profiles in the S1 and
the S3 cases (S2 is very similar to S1). The shift profiles are
modulated by the lapse ones, so that the shift goes to zero in the
collapsed regions. This is a consequence of the term $-\alpha\,
Q\,\beta^i$ in the shift evolution equation, devised for getting
finite values of the combination $\beta^i/\alpha$. In the
non-collapsed region, S1 leads to a higher shift profile, which
spreads out with time, whereas S3 leads to a lower profile, which
starts diminishing after the initial growing. Allowing for
(\ref{S3}), this indicates that the conformal gamma quantity
$\widetilde{\Gamma}_i$ is driven to zero. The lapse slopes are
also slightly softened in the S3 case.

These results confirm that the code stability is not linked to any
particular shift prescription, as we can combine different source
terms in the shift evolution equation, leading to different lapse
and shift profiles.

\section{Conclusions and Outlook}
We have shown in this paper how a first-order flux-conservative
version of the Z4 formalism can be adjusted for dealing with the
ordering constraints, and then implemented in a numerical code by
means of a robust, cost-efficient, finite-difference formula. The
resulting scheme has been tested in a demanding
harmonic-coordinates scenario: the gauge-waves testbed. The code
performance compares well with the best harmonic-code results for
this test~\cite{Mexico2}, even in the highly non-linear regime
($50\%$ amplitude case). This is in contrast with the well-known
problems of BSSN-based codes with the gauge-waves
test~\cite{KS08}~\cite{Mexico}.

The code has also been tested in non-excision BH evolutions, where
singularity-avoidance is a requirement. Our results confirm the
robustness of the code for many different choices of dynamical
lapse and shift prescriptions. In the normal coordinates case
(zero shift), our results set up a new benchmark, by evolving the
BH up to $1000\,M$ without any sign of instability. This improves
the reported BSSN result by one order of magnitude (Harmonic codes
are not devised for normal coordinates). More important, this
shows that a specific shift choice is not crucial for code
stability, even in non-excision BH simulations. This is confirmed
by our shift simulations, where different covariant evolution
equations for the shift lead also to stable numerical evolution.

In spite of the encouraging performance in these basic tests, we
still are on the way towards a gauge-polyvalent code, as pointed
out by the title of this paper. More technical developments on the
numerical part are required: mesh refinement, improved boundary
treatment, etc. On the theoretical side, as far as the shift
prescription is no longer determined by numerical stability, we
can explore shift choices from the physical point of view,
adapting our space coordinates system to the features of every
particular problem. We are currently working in these directions.

\section*{Acknowledgments}
This work has been jointly supported by European Union FEDER
funds, the Spanish Ministry of Science and Education (projects
FPA2007-60220, CSD2007-00042 and ECI2007-29029-E) and by the
Balearic Conselleria d'Economia Hissenda i Innovaci\'{o} (project
PRDIB-2005GC2-06). D.~Alic and C.~Bona-Casas acknowledge the
support of the Spanish Ministry of Science, under the
BES-2005-10633 and FPU/2006-02226 fellowships, respectively

 ~
\renewcommand{\theequation}{A.\arabic{equation}}
\setcounter{equation}{0}
\section*{Appendix A: Flux-Conservative evolution equations}
We will write the first-order evolution system in a balance-law
form. For a generic quantity $u$, this leads to
\begin{equation}\label{balance}
    \partial_t~u + \partial_k~F^k(u) = S(u)~,
\end{equation}
where the Flux $F^k(u)$ and Source terms $S(u)$ can depend on the
full set of dynamical fields in an algebraic way. In the case of
the space-derivatives fields, their evolution equations
(\ref{dtA1}-\ref{dtD1}) are yet in the balance-law form
(\ref{balance}). Note however that any damping terms of the form
described in (\ref{Adamped}) will contribute both to the Flux and
the Source terms in a simple way.

The metric evolution equation (\ref{dtgamma}) will be written in
the form
\begin{equation}
    \partial_t~ \gamma_{ij} = 2\,\beta^k D_{kij} + B_{ij} + B_{j\,i}
   - {2~\alpha}~K_{ij}~,
\end{equation}
so that it is free of any Flux terms. The remaining (non-trivial)
evolution equations (\ref{dtK}- \ref{dtZ}) require a more detailed
development. We will expand first the Flux terms in the following
way:
\begin{eqnarray}
\partial_t K_{ij} &+&  \partial_k [- \beta^k~K_{ij} + \alpha\; {\lambda^k}_{ij}~] =
S(K_{ij})
\\
\partial_t Z_i &+& \partial_k [-\beta^k~Z_i + \alpha~
 \{-{K^k}_i + {\delta^k}_i (trK - \Theta) \}
 \\
 &~&\qquad +\mu~ ({B_i}^k-{\delta_i}^k trB)~]
 = S(Z_i) \nonumber
 \\
\partial_t \Theta &+&  \partial_k [- \beta^k~\Theta + \alpha\;(D^k - E^k - Z^k)~]
 = S(\Theta)
\end{eqnarray}
where we have used the shortcuts $D_i\equiv {D_{ik}}^k$ and $E_i
\equiv {D^k}_{ki}$~, and
\begin{eqnarray}\label{flux_K}
    {\lambda^k}_{ij} = {D^k}_{ij}
     - \frac{1}{2}~(1 + \xi)\; ({D_{ij}}^k + {D_{j\,i}}^k)
    &+& \frac{1}{2}~ {\delta^k}_i\; [A_j + D_j - (1 - \xi)\; E_j -2~Z_j\,] \\
    &+& \frac{1}{2}~ {\delta^k}_j\; [A_i + D_i - (1 - \xi)\; E_i
    -2~Z_i\,]\,.
    \nonumber
  \end{eqnarray}

The Source terms $S(u)$ do not belong to the principal part and
will be displayed later. Let us focus for the moment in the
hyperbolicity analysis, by selecting a specific space direction
$\vec{n}$, so that the corresponding characteristic matrix is
\begin{equation}\label{Amatrix}
    {A}^n =
    \frac{\partial\,{F}^n}{\partial\,{u}}\,,
\end{equation}
where the symbol $n$ replacing an index stands for the projection
along the selected direction $\vec{n}$. We can get by inspection
the following (partial) set of eigenfields, independently of the
gauge choice:
\begin{itemize}
    \item \textbf{Transverse derivatives}:
    \begin{equation}\label{transverse}
    A_\bot\,, \qquad {B_\bot\,}^i\,,\qquad
    D_{\bot ij}\,,
\end{equation}
propagating along the normal lines (characteristic speed
$-\beta^n\,$). The symbol $\bot$ replacing an index means the
projection orthogonal to $\vec{n}$.
    \item \textbf{Light-cone eigenfields}, given by the pairs
    \begin{eqnarray}\label{Wtransv}
      F^n[\,D_{n\bot\bot}\,] &\pm&  F^n[\,K_{\bot\bot}\,]\\
      \label{Wmixed}
       -F^n[\,Z_{\bot}\,] &\pm& F^n[\,K_{n\bot}\,] \\ \label{Wenergy}
      F^n[\,D_{n}-E_n-Z_n\,] &\pm& F^n[\,\Theta\,]
    \end{eqnarray}
    with characteristic speed $-\beta^n \pm \alpha\,$,
    respectively.
\end{itemize}

Note that the eigenvector expressions given above, in terms of the
Fluxes, are valid for any choice of the ordering parameters $\mu$
and $\xi$. Only the detailed expression of the eigenvectors,
obtained from the Flux definitions, is affected by these parameter
choices. For instance
\begin{equation}\label{F_Vn}
    F^n[\,D_{n}-E_n-Z_n\,] = -\beta^n\, [\,D_{n}-E_n-Z_n\,]
    + \alpha\,\theta + (\mu-1)\,tr(B_{\bot\bot})\,.
\end{equation}
Any value $\mu\neq 1$ implies that the characteristic matrix of
the Z3 system, obtained by removing the variable $\theta$ from our
Z4 evolution system~\cite{Z48}, can not be fully diagonalized in
the dynamical shift case. Of course, the hyperbolicity analysis
can not be completed until we get suitable coordinate conditions,
amounting to some prescription for the lapse and shift sources $Q$
and $Q_i$, respectively. But the subset of eigenvectors given here
is gauge independent: non-diagonal blocs can not be fixed {\em a
posteriori\,} by the coordinates choice.

The detailed expressions for the eigenvectors can be relevant when
trying to compare with related formulations. For instance, a
straightforward calculation shows that the eigenvectors
(\ref{Wtransv}-\ref{Wenergy}) can be matched to the corresponding
ones in the harmonic formalism if and only if
\begin{equation}\label{mu_xi}
    \xi=-1\,,\qquad \mu=1/2\,.
\end{equation}
This shows that different requirements can point to different
choices of these ordering parameters. We prefer then to leave this
choice open for future applications. Concerning the simulations in
this paper, we have taken $\xi=-1\,$, $\mu=1\,$.

Finally, we give for completeness the Source terms, namely:
\begin{eqnarray}\label{sources}
    S(K_{ij}) &=& - K_{ij}~trB + K_{ik}~{B_j}^k +
    K_{jk}~{B_i}^k + \alpha\; \{\frac{1}{2}\; (1 + \xi)\; [-A_k\; {\Gamma^k}_{ij}
      + \frac{1}{2}(A_i\;D_j + A_j\;D_i)] \nonumber \\
      &+& \frac{1}{2}\; (1 - \xi)\; [A_k\; {D^k}_{ij}
    - \frac{1}{2} \{A_j\; (2\; E_i - D_i) + A_i\; (2\; E_j - D_j)  \}
    \nonumber \\
    &+& 2\; ({D_{ir}}^m ~ {D^{r}}_{mj}  + {D_{jr}}^m ~ {D^{r}}_{mi})
    - 2\; E_k\; ({D_{ij}}^k + {D_{ji}}^k)] \nonumber  \\
    &+& (D_k + A_k - 2\; Z_k) \; {\Gamma^k}_{ij}
    - {\Gamma^k}_{mj}\; {\Gamma^m}_{ki} - (A_i\; Z_j + A_j\; Z_i)
    - 2\; {K^k}_i\; K_{kj} \nonumber \\
    &+& (trK - 2\; \Theta)\; K_{ij}\}
    - 8\; \pi\; \alpha\; [S_{ij} - \frac{1}{2}~(trS - \tau)~\gamma_{ij}]
\\
   S(Z_i) &=& - Z_{i}~trB + Z_{k}~{B_i}^k
   + \alpha\; [A_i\; (trK - 2\; \Theta) - A_k\; {K^k}_i
   - {K^k}_r\; {\Gamma^r}_{ki} + {K^k}_i\; (D_k - 2\; Z_k)]
   \nonumber
\\
   &~&\qquad - 8\; \pi\; \alpha\; S_i
\\
   S(\Theta) &=& -\Theta~trB
      + \frac{\alpha}{2}  [2\; A_k\; (D^k - E^k - 2\; Z^k)
      + {D_k}^{rs}\; {\Gamma^k}_{rs} \nonumber - D^k (D_k - 2\;
      Z_k) - {K^k}_r\; {K^r}_k \nonumber \\
      &+& trK\; (trK - 2\; \Theta)] - 8\; \pi\; \alpha\; \tau ~.
\end{eqnarray}

\renewcommand{\theequation}{B.\arabic{equation}}
\setcounter{equation}{0}
\section*{Appendix B: Finite-differences implementation}

We follow the well-known method-of-lines (MoL~\cite{MoL}) in order
to deal separately with the space and the time discretization.
Concerning the time discretization, we use the following
third-order-accurate Runge-Kutta algorithm
\begin{eqnarray}\label{RK3}\nonumber
    u^* &=& u^n + \Delta t~rhs(~u^n~)\\
    u^{**} &=& \frac{3}{4}~u^n + \frac{1}{4}~[u^* + \Delta t~rhs(~u^*~)]\\
    u^{n+1} &=& \frac{1}{3}~u^n + \frac{2}{3}~[u^{**}+\Delta t~rhs(~u^{**}~)]~,
\nonumber
\end{eqnarray}
which is strong-stability-preserving (SSP~\cite{SSP}), where we
have used as a shorthand
\begin{equation}\label{rhs}
    rhs(~u~) \equiv  -\,\partial_k\,F^k(u) + S(u)~.
\end{equation}

The Flux derivatives appearing in (\ref{rhs}) will be discretized
by using the finite-difference formula proposed in
Ref.~\cite{JCPpaper} (FDOC algorithm). For instance the derivative
of $F^x(u)$ will be represented as
\begin{equation}\label{DFform}
    \partial_x F^x_j = C^{2m}F^x_j + (-1)^{m}\beta (\Delta x)^{2m}
        D_{+}^{m}D_{-}^{m-1}(\lambda_{j-1/2}\,D_{-}u_j)~,
\end{equation}
where $C^{2m}$ is the $2m$th-order-accurate central difference
operator and $D_\pm$ are the standard finite difference operators.
We have also noted
\begin{equation}\label{lambda}
    \lambda_{j-1/2} = max(\lambda_j,\lambda_{j-1})~,
\end{equation}
where $\lambda_j$ stands here for the local characteristic radius
(the highest characteristic speed, typically the gauge speed).

Note that the second term in the finite-difference formula
(\ref{DFform}) is actually a dissipation operator of order $2m$
acting on ($\lambda \,u$), so it could be regarded at the first
sight as a mere generalization of the standard Kreiss-Oliger
artificial viscosity operators~\cite{KO}. This is not the case:
the formula (\ref{DFform}) can be instead derived in a
finite-volume framework, when combining the local-Lax-Friedrichs
flux formula~\cite{Rusanov} with the (unlimited)
Osher-Chakrabarthy flux interpolation~\cite{ICASE} (see
Ref.~\cite{JCPpaper} for details, including the optimal values of
the $\beta$ parameter).

Note that, contrary to the standard Kreiss-Oliger approach, the
dissipation term is such that the accuracy of the first (centered
derivatives) term in (\ref{DFform}) is reduced by one order: the
resulting FDOC algorithm accuracy is always of an odd order. This
is important for code robustness. The algorithms (\ref{DFform})
can be shown to keep monotonicity even for remarkably high
compression factors (defined as the ratio between two neighbor
slopes along a given direction)~\cite{JCPpaper}, which is what is
actually required in view of the steep profiles shown for instance
in Fig.~\ref{BHshort}.

The space accuracy of the scheme (\ref{DFform}) is $2m-1$, with an
stencil of $2m+1$ points. We have used in this paper both the
third-order and fifth-order accurate methods, for which the
optimal values of the dissipation parameter are $\beta=1/12$,
$\beta=2/75$, respectively~\cite{JCPpaper}. In the fifth order
case, we have a seven-point stencil and the dissipation term
corresponds to a sixth derivative, as in the advanced
finite-difference schemes used in Ref.~\cite{Jena08}. The
robustness of the proposed algorithms, with compression factors of
$5$ and $3$ respectively, makes them very convenient for
steep-gradient scenarios, such us the ones arising in black-hole
simulations, where slice-stretching threatens the stability of
more standard finite-difference algorithms~\cite{stretching}.

No sophisticated numerical tools (mesh refinement,
algorithm-switching for the advection terms, etc) have been
incorporated to our code at this point, when we are facing just
test simulations. Concerning the boundary treatment, we simply
choose at the points next to the boundary the most accurate
centered algorithm compatible with the available stencil there.
When it comes to the last point, we can either copy the neighbor
value or propagate it out with the maximum propagation speed (by
means of a 1D advection equation). The idea is to keep the
numerical code as simple as possible in order to test here just
the basic algorithm in a clean way.

\renewcommand{\theequation}{C.\arabic{equation}}
\setcounter{equation}{0}
\section*{Appendix C: Scalar field stuffing}
Let us consider the stress-energy tensor
\begin{equation}\label{Tab}
    T_{ab} = \Phi_a~\Phi_b - 1/2\,(g^{cd}\Phi_c~\Phi_d)~g_{ab}~,
\end{equation}
where we have noted $\Phi_a=\partial_a\,\Phi$, corresponding to a
scalar field matter content. The 3+1 decomposition of (\ref{Tab})
is given by
\begin{equation}\label{Sij}
    \tau = 1/2\,({\Phi_n}^2+\gamma^{kl}\Phi_k\,\Phi_l)\,,\qquad
    S_i = \Phi_n\,\Phi_i\,,\qquad
    S_{ij} = \Phi_i\,\Phi_j
    +1/2\,({\Phi_n}^2-\gamma^{kl}\Phi_k\,\Phi_l)\,\gamma_{ij}~,
\end{equation}
where $\Phi_n$ stands for the normal time derivative:
\begin{equation}\label{Phin}
    (\partial_t -\beta^k\,\partial_k)~ \Phi = - \alpha~\Phi_n\,.
\end{equation}
The quantities (\ref{Sij}) appear as source terms in the field
equations (\ref{dtK}-\ref{dtZ}).

The stress-energy conservation amounts to the evolution equation
for the scalar field, which is just the scalar wave equation. In
the 3+1 language, it translates into the Flux-conservative form:
\begin{equation}\label{dtPhin}
    \partial_t~[\,\sqrt{\gamma}~\Phi_n\,] + \partial_k~[\,\sqrt{\gamma}~
    (-\beta^k \Phi_n + \alpha\,\gamma^{kj} \Phi_j)\,] = 0\,.
\end{equation}
A fully first-order system may be obtained by considering the
space derivatives $\Phi_i$ as independent dynamical fields, as we
did for the metric space derivatives.

Concerning the initial data, we must solve the energy-momentum
constraints. They can be obtained by setting both $\Theta$ and
$Z_i$ to zero in (\ref{dtTheta}, \ref{dtZ}). In the time-symmetric
case ($K_{ij}=0$), this amounts to
\begin{equation}\label{EMconstraints}
    R = 16\pi\,\tau\,,\qquad S_i = \Phi_n\,\Phi_i = 0\,.
\end{equation}
The momentum constraint will be satisfied by taking $\Phi$ (and
then $\Phi_i$) to be zero everywhere on the initial time slice.
Concerning the energy constraint, we will consider the line
element (\ref{isotropic}) with $m=m(r)$. We assume a constant mass
value $m=M$ for the black-hole exterior, so that the energy
constraint in (\ref{EMconstraints}) will be satisfied with
$\tau=0$ there.

In the interior region, the energy constraint will translate
instead into the equation
\begin{equation}\label{meq}
    m'' = - 2\pi r~(\Phi_n)^2~(1+\frac{m}{2r})^5~,
\end{equation}
which can be interpreted as providing the initial $\Phi_n$ value
for any convex ($m''\leq 0$) mass profile. Of course, some
regularity conditions both at the center and at the matching point
$r_0$ must be assumed. Allowing for (\ref{meq}), we have taken
\begin{eqnarray}
\nonumber
  m = m'' &=& 0\qquad (r=0) \\
\nonumber  m=M,\qquad m'=m'' &=& 0\qquad (r=r_0)~.
\end{eqnarray}

Note that, allowing for (\ref{meq}), these matching conditions
ensure just the continuity of $\Phi_n\,$, not its smoothness. This
can cause some numerical error, as we are currently evolving
$\Phi_n$ through the differential equation (\ref{dtPhin}). If this
is a problem, we can demand the vanishing of additional
derivatives of the mass function $m(r)$, both at the origin and at
the matching point (this is actually the case in our shift
simulations). This is not required in the standard case
($f=2/\alpha$, normal coordinates), where we have used a simple
profile, with the matching point at the apparent horizon
($r_0=M/2$), given by
\begin{equation}\label{mprofile}
    m(r) = 4r - 4/M [\,r^2 + (M/2\pi)^2~sin^2(2\pi r/M)\,]~.
\end{equation}

\bibliographystyle{prsty}

\end{document}